\documentclass[10pt]{article}
\usepackage{graphics}
\usepackage{epsfig}
\usepackage{flafter}
\usepackage{latexsym}
\begin{document}
%\setcounter{chapter}{5}
%\include{Titel}
%\tableofcontents
\bibliographystyle{plain}

\begin{center}{\Large\bf Weighing nearby stars with GAIA?}\end{center}

\vspace{0.2cm}

\begin{center} S. Dib\raisebox{1ex}{\scriptsize 1}, J. Surdej\raisebox{1ex}{\scriptsize 2,}\raisebox{1ex}{\scriptsize $\ast$}, J.-F. Claeskens\raisebox{1ex}{\scriptsize 2,} \raisebox{1ex}{\scriptsize $\ast\ast$} \end{center} 

\vspace{0cm}

\hspace*{0cm} \raisebox{1ex }{\scriptsize 1}{\footnotesize  Max Planck Institut f\"ur Astronomie, K\"onigstuhl 17, 69117 Heidelberg, Germany}\\
\hspace*{0cm} \raisebox{1ex}{\scriptsize 2}{\footnotesize Institut d'Astrophysique et de G\'{e}ophysique, 5 Avenue de Cointe, 4000 Li\`{e}ge, Belgium} \newline
{\footnotesize \raisebox {1ex}{\scriptsize $\ast$} Directeur de Recherches du FNRS (Belgium)} \newline
{\footnotesize \raisebox  {1ex}{\scriptsize $\ast\ast$} Charg\'e de Recherches du FNRS (Belgium)} \newline
\vspace{0.25cm}
 
\begin{center}{\Large {Abstract}}\end{center}

 Microlensing consists in two major effects: (1) variation in the apparent position of the background sources (astrometric component) and (2) flux variations of the background sources (photometric component). While the latter has been extensively used in the search for dark objects in the Galactic disk and halo (projects like MACHO (Alcock {\it et al}, 1997), EROS (Derue {\it et al}, 1999), OGLE (Paczy$\acute{n}$ski {\it et al}, 1994)), the first effect has not yet been part of a systematic observational program, simply because the observations of very slight displacements in the positions of background sources requires an astrometric accuracy which current telescopes do not yet provide. We investigate here whether the astrometric accuracy of GAIA could enable such measurements and, as a consequence, enable new, direct and original measurements of the mass of nearby stars. 

\vspace{0.3cm}
{\Large 1.   Introduction}
\vspace{0cm}
 
\begin{figure}[h] 
\includegraphics[width=10cm]{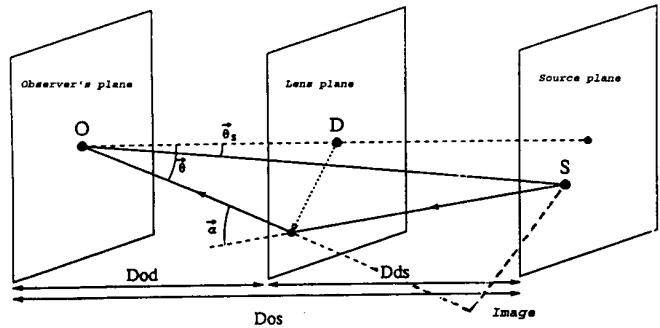}
\centering
\caption{\small Schematic geometrical configuration of a microlensing system.}
\label{figune}
\end{figure}

When a nearby star (the lens) passes in the foreground of a distant background source, the luminous ray coming from the source is deflected in the vicinity of the lens by it's gravitationnal potential, and two distinct images of the source are produced, only one of which (called the main image) is amplified and can be observed. A schematic description of a microlensing system is given in Figure \ref{figune}, where O represents the observer, D the lens and S the background source.
\noindent
A simple equation, named the lens equation links the position of the source $\theta_{s}$, the position of the image $\theta$  and the deflection angle $\alpha$. When described by the simple Point Mass model; See Resfdal and Surdej (1994), the lens equation is a simple second order equation :
\begin{eqnarray} 
\theta-\theta_{S} &  = &  \frac {4\;G\;M} {c^{2}} \; \frac 1 \theta \; \frac  {D_{DS}} {D_{OS}\;D_{OD}},   
\label{eq1}     
\end{eqnarray}          
\begin{figure}[h]
\centering
\includegraphics[width=9.5cm]{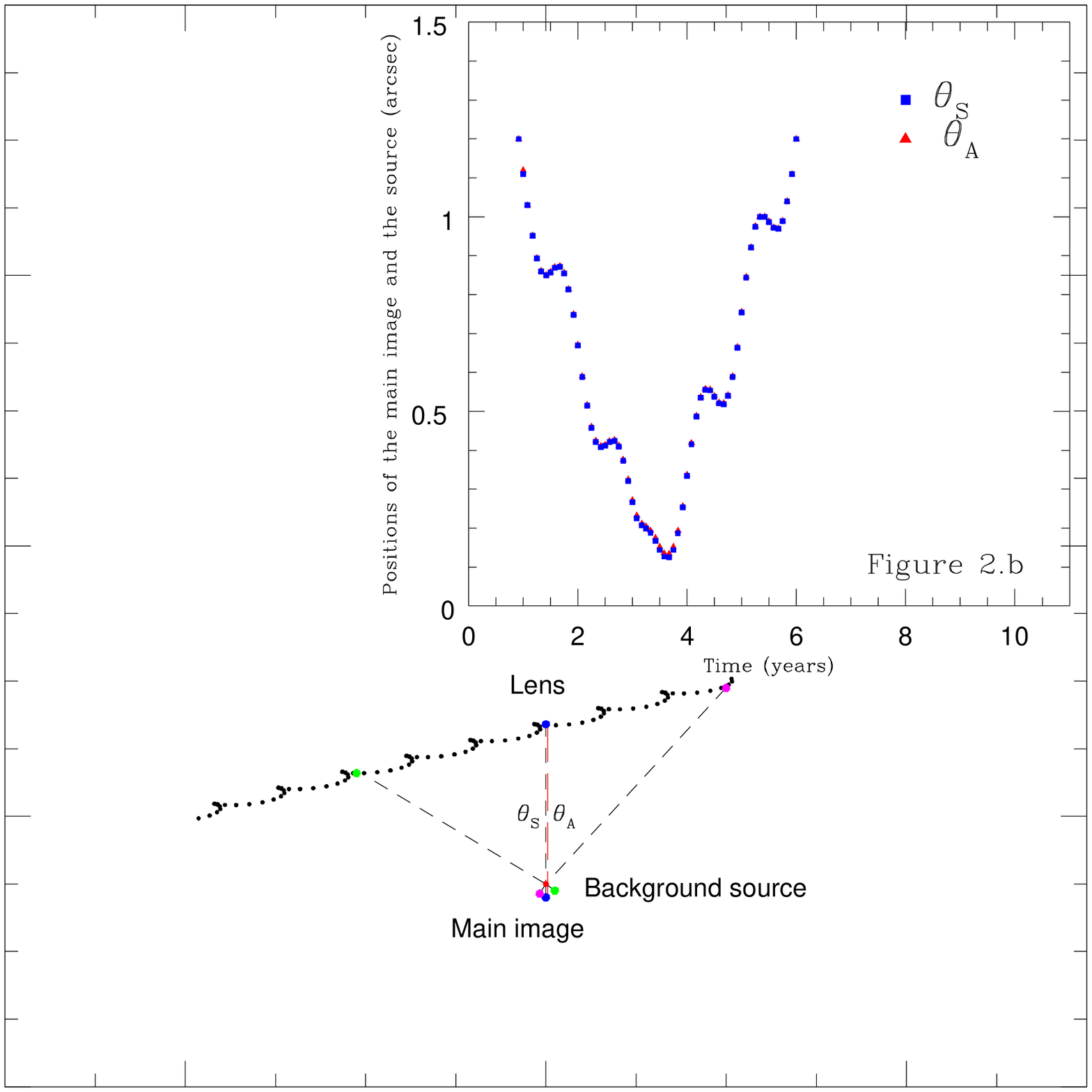}
\caption{\small a) Moving microlens and b) Astrometric  microlensing effects.}
\label{figdeux}
\end{figure}        
\noindent
where $D_{DS}$, $D_{OS}$ and $D_{OD}$ are respectively the lens-source, observer-source and observer-lens distances and M the mass of the lens. The right hand term of Equation \ref{eq1} stands for the deflection angle $\alpha$. For a nearby star the ratio between  $D_{DS}$ and  $D_{OS}$ is close to unity and the Einstein angular radius simplifies to:
\begin{eqnarray}
\theta_E \cong  \sqrt{\frac {4\;G\;M} {D_{OD}\;c^2} }.               
\end{eqnarray}
\noindent                 
The positive root of Equation \ref{eq1} gives the position of the main image $\theta_A$. The difference between  $\theta_A$ and $\theta_S$ is the signature of the microlensing effect. In order to quantify this effect, we artificially place a 1.5 M\raisebox{-1ex}{\scriptsize $\odot$} lens, which has a proper motion of 0.4"/year, at a distance of 10 parsec and a background source at an initial angular separation of 1.2 arcsec from the lens. In Figure \ref{figdeux}.a we display the trajectory of the lens projected on the sky, the background source (red dot) and the main image which is formed due to microlensing. Whenever the difference between the position of the source $\theta_{S}$ and the position of the main image $\theta_{A}$ can be resolved (here of the order $10^{-3}$-$10^{-2}$ arcsec), a lensing event is registered and astrometric lensing effects measured.
%\begin{figure} [h]
%\centering 

\begin{figure} [h]
\setlength{\unitlength}{1cm}
\begin{minipage}[t]{6.2cm} 
\begin{picture} (6.2,6.2) \includegraphics[width=6.2cm]{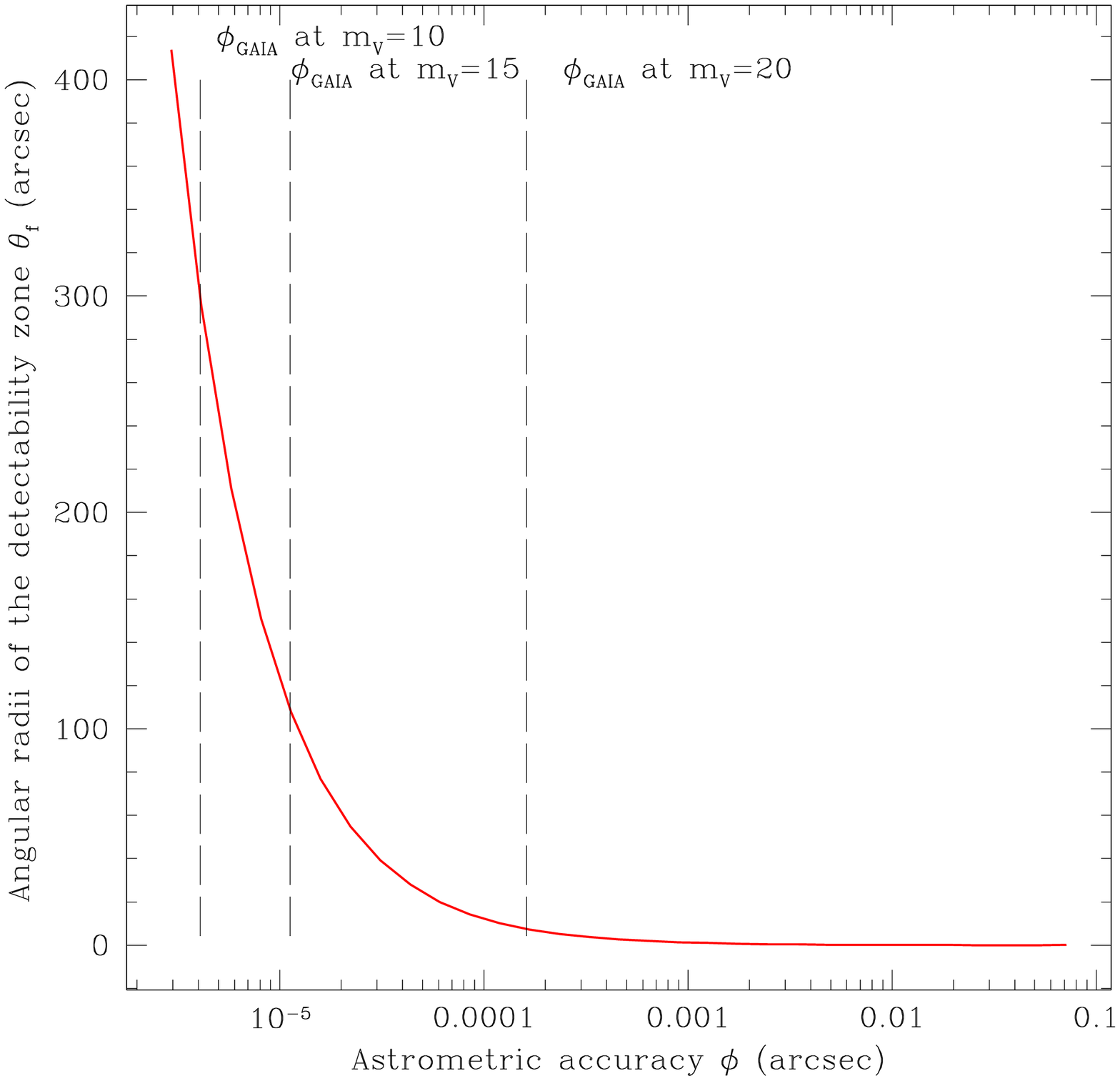} \end{picture}\par
\caption{\small Angular radii of the astrometric detection zone as a function of the astrometric accuracy.}
\label{figtrois}
\end{minipage}\hfill
\begin{minipage}[t] {6.2cm}
\begin{picture} (6.2,6.2) \includegraphics[width=6.2cm]{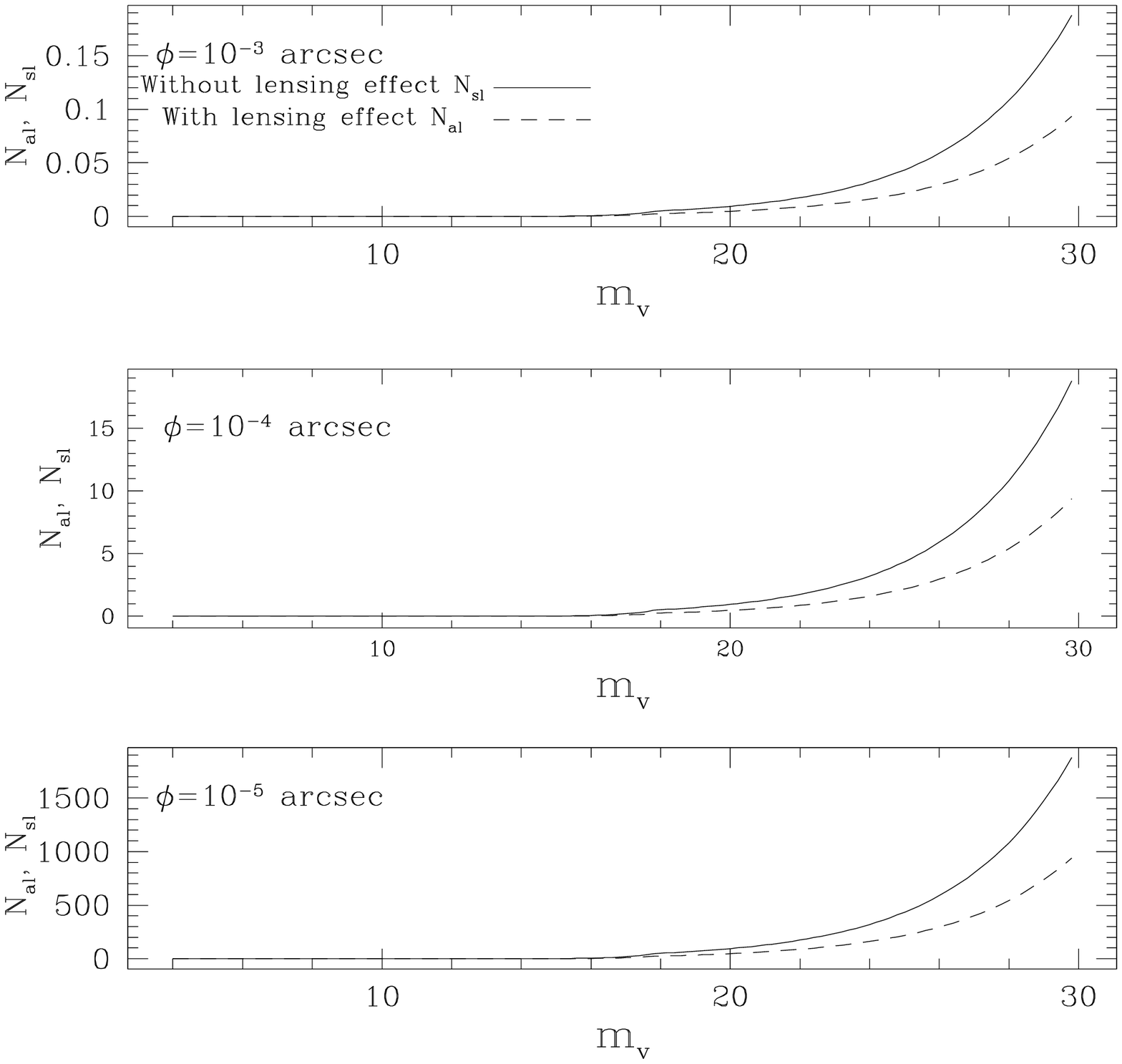} \end{picture}\par
\caption{\small Number of stars in the detection zone as a function of $\phi$ and $m_{v}$ if microlensing effects are $(N_{al})$ or not $(N_{sl})$ taken into account.} 
\label{figquatre}
\end{minipage}
\end{figure}

The physical quantity that is extracted from the fitting of the positions of the lensed background sources as displayed in Figure \ref{figdeux}.b is not directly the mass of the lens but its Einstein angular radius $\theta_{E}$ which does not only depend on the mass of the lens  but also on its distance. Thus, large uncertainties on the distance will control those on the mass. Another point is that the lens must have a proper motion high enough,  typically of the order $\simeq$ (0.3"-1"/year), so that its detection zone (see below) will cover a part of the sky sufficiently large during a time span that is short enough for observations to be carried out. This is why the method is ideally suited for weighing nearby stars.  

\vspace{0.2cm} 
{\Large 2.   Application of the method to GAIA}
\vspace{0cm}

A lensing event is detectable whenever  $\theta_{A}(t)-\theta_{S}(t)\ge\phi$, where $\phi$ represents an astrometric accuracy. This leads to the concept of a circularly shaped detection zone centered on the lens. Figure \ref{figtrois} displays the radii of the detection zone $\theta_{f}$ for the above considered lens as a function of the astrometric accuracy. The three vertical dashed lines in Figure \ref{figtrois} correspond to three different expected accuracies for GAIA at three different limiting magnitudes, see Lindegreen and Perryman (1996). The ratio of the area of the astrometric detection zone to the area of the photometric detection zone at the magnitudes of 10, 15 and 20 is given by $\theta_{f}(m_{v}=10,15,20)/\theta_{E}$ which yields respectively the values of $7.5\times10^{7}$, $1\times10^{7}$ and $5\times10^{4}$. Based on the Bahcall \&  Soneira star counting models (1981), we calculate for a given astrometric accuracy $\phi$ the number of source stars in the background of the lens (In the direction of the Galactic center, i.e. the optimal direction to detect microlensing events). The number of stars in the detection zone when microlensing effects are or not taken into account are represented by $N_{al}$ and $N_{sl}$, respectively. $N_{al}$ and $N_{sl}$ are calculated as a function of the limiting magnitude $m_{v}$. This is displayed in Figure \ref{figquatre} for three different values of the astrometric accuracy. Table \ref{tabune} summarizes the expected number of source stars in the background of the lens for the three specific GAIA accuracies indicated in Figure \ref{figtrois} along with the value of the corresponding detection zone radii. A minimum criterion for the applicability of this method is to have at least one background source in the detection zone of the lens (under the condition that the detection zone might be crossed by the lens during a span of time short enough to carry out the observations). Ideally, more than one background source should be present in the detection zone to allow for a statistical treatment and to reduce uncertainties. The numbers $N_{al}$ displayed in Table 1 are quite small although the corresponding detection zones are quite large. This compromises the observation of complete curves such as the one displayed in Figure \ref{figdeux}.b. Ideally, if an astrometric accuracy of 4 ${\mu}as$ was available at the limiting magnitude $m_{v}=14$, the number $N_{al}$ would be equal to 2 and equal to 17 at $m_{v}=16$.

\vspace{0.3cm}
\begin{table} [ht]
\begin{center}
\begin{tabular}[c]{|l|l|l|c|} \hline
  & $N_{sl}$ & $N_{al}$ & $\theta_{f}$ (arcsec)\\ \hline\hline
$\phi=4\;{\mu}as$, $m_{v}=10$  & $4.62\times10^{-2}$  &  $2.03\times10^{-2}$   & 305 \\ \hline
$\phi=11\;{\mu}as$, $m_{v}=15$  & 1.68  &   0.84  & 111 \\ \hline
$\phi=160\;{\mu}as$, $m_{v}=20$  & 0.37  &  0.185  & 7.63 \\ \hline
\end{tabular}
\caption{\small Summary of the expected number of stars in the detection zone of the test star for three specific GAIA accuracies}
\label{tabune}
\end{center}
\end{table}
\vspace{0.2cm}
{\Large 3. Conclusion}
\vspace{0cm}

Due to the low background surface density of stars at optical wavelengths, GAIA presents too limited capabilities in order to detect microlensing events that are suitable for weighing nearby stars. Better astrometric accuracies of future space missions as well as operating these at longer wavelenghts could provide the necessary improvements to achieve this goal. 

\vspace{0.1cm}
{\Large References}

{\small Alcock C. {\it et al} 1997, ApJ. vol. 491, 436}\newline
{\small Bahcall J. N. \& Soneira R. M. 1980, ApJ. Suppl. Ser. vol. 44, 73}\newline
{\small Derue {\it et al} 1999, EROS collaboration A\&A vol. 351, 87}\newline
{\small Lindegreen L. \& Perryman M. A. C. 1996, A\&A Suppl. Ser. vol. 116, 579}\newline
{\small Paczy$\acute{n}$ski B. {\it et al} 1994, ApJ. Part2-Letters vol. 435, nr2 L113}\newline
{\small Resfdal S. \& Surdej 1994, J. Rep. Prog. Phys. vol. 56, 117}  
\end{document}